\documentclass[journal=nalefd,manuscript=letter]{achemso}

\usepackage{graphicx}
\usepackage{cleveref}
\usepackage[hidelinks]{hyperref}
\usepackage{color}
\usepackage{amsmath,amssymb,amstext,graphicx,bm}

\author{Lada~Vuku\v{s}i\'c}
\email{lada.vukusic@ist.ac.at}
\phone{+43 2243 9000 721204}
\affiliation[IST Austria]{Institute of Science and Technology Austria, Am Campus 1, 3400 Klosterneuburg, Austria}

\author{Josip~Kuku\v{c}ka}
\email{josip.kukucka@ist.ac.at}
\affiliation[IST Austria]{Institute of Science and Technology Austria, Am Campus 1, 3400 Klosterneuburg, Austria}

\author{Hannes~Watzinger}
\affiliation[IST Austria]{Institute of Science and Technology Austria, Am Campus 1, 3400 Klosterneuburg, Austria}

\author{Georgios~Katsaros}
\affiliation[IST Austria]{Institute of Science and Technology Austria, Am Campus 1, 3400 Klosterneuburg, Austria}
\alsoaffiliation[JK University]{Johannes Kepler University, Institute of Semiconductor and Solid State Physics, Altenbergerstr. 69, 4040 Linz, Austria}

\title{Fast hole tunneling times in Germanium hut wires probed by single-shot reflectometry}

\begin{document}

\begin{abstract}
{Heavy holes confined in quantum dots are predicted to be promising candidates for the realization of spin qubits with long coherence times. Here we focus on such heavy-hole states confined in Germanium hut wires. By tuning the growth density of the latter we can realize a T-like structure between two neighboring wires. Such a structure allows the realization of a charge sensor, which is electrostatically and tunnel coupled to a quantum dot, with charge-transfer signals as high as 0.3\,e. By integrating the T-like structure into a radio-frequency reflectometry setup, single-shot measurements allowing the extraction of hole tunneling times are performed. The extracted tunneling times of less than 10\,$\mu$s are attributed to the small effective mass of Ge heavy-hole states and pave the way towards projective spin readout measurements.}
\end{abstract}

Keywords: Germanium, quantum dot, heavy hole, reflectometry, single-shot measurement

\newpage

Spin qubits realized in p-type group IV materials \cite{Maurand2016} have emerged as an alternative to electron-based qubit systems \cite{Hanson2007, ZwanenburgReview2013}. They have the advantage of lower hyperfine interaction, leading to long dephasing times even in natural samples \cite{HigginbothamNanoLett2014}, and short manipulation times due to the strong spin orbit coupling \cite{Maurand2016, HaoNanoLett2010, KloeffelPRB2011, HigginbothamPRL2014, LiNanoLett2015}. In particular, long spin lifetimes are predicted when the confined states are of heavy-hole (HH) character \cite{BulaevLoss2005,Fisher2008}. Such states have been recently achieved in Si quantum dots (QDs) \cite{LiNanoLett2015} as well as for holes confined in so-called Ge hut wires (HWs) \cite{Watzinger2016}.

For any qubit experiment, the realization of a high fidelity spin readout scheme is essential. In the initial experiments, a quantum point contact capacitively coupled to a QD hosting the qubit was used for the spin-to-charge conversion \cite{Elzerman2004}. Later on, it was demonstrated that a capacitively coupled QD could also act as a sensitive electrometer \cite{Hu2007}. In 2009, Morello et al.\cite{MorelloArchitecture} suggested to use a charge sensor which is not only capacitively but also tunnel coupled to the spin qubit. Such a structure led to high charge-transfer signals, opening the path for fast and high-fidelity single-shot readout measurements \cite{MorelloNature2010, Muhonen2014, Watson2015}.

Here we report on the realization of a charge sensor for a p-type QD formed in a Ge HW \cite{Zhang2012, Watzinger2014}. Low-temperature transport measurements reveal charge-transfer signals as large as 0.3\,\textit{e} visible already at 1.5\,K. By incorporating the QD charge sensor device in a radio-frequency (RF) reflectometry setup \cite{Schoelkopf1998, ReillyAPL07, PettaAPL12, AresPRA16}, single hole tunneling events can be observed. Single-shot RF reflectometry measurements reveal tunneling times between the QD and the charge sensor shorter than 10\,$\mu$s. These short tunneling times are attributed to the relatively small effective mass of Ge HHs when transport takes place in the growth plane.    

\begin{figure}
\includegraphics{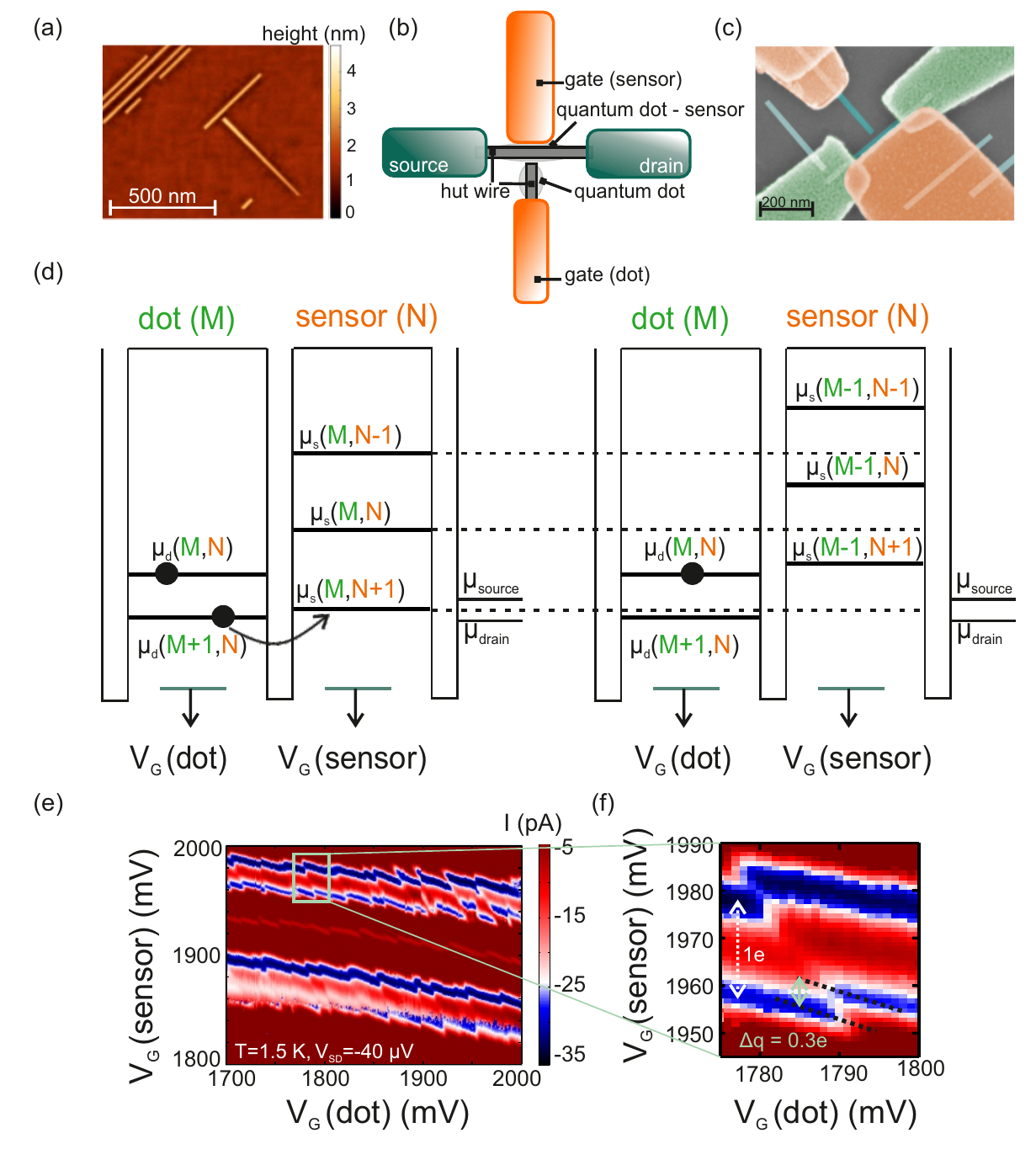}
\caption{\label{fig:figure1} (a) Atomic force microscopy image showing a T-like Ge HW structure. (b) Schematic of the charge sensor device used in this work. The two perpendicular HWs are shown in dark gray and the estimated position of the formed QDs in the wires in light gray. The source and drain of the sensor - SHT - are shown in dark green and the gates of the sensor and the QD are shown in orange. (c) False color scanning electron micrograph of a device similar to those measured. (d) Schematic showing the ladder of electrochemical potentials of a capacitive and tunnel coupled QD-sensor system for the two cases of the dot occupancy, M and M+1. (e) Stability diagram obtained by sweeping the gate of the QD versus the gate of the charge sensor, at the source-drain (V$_{\textrm{SD}}$) bias of -40\,$\mu$V. Every time when the number of holes in the dot changes, the Coulomb peak of the SHT breaks and shifts. (f) Zoom-in into the stability diagram, showing the discontinuity of a Coulomb peak. The dashed lines with the green (solid) double arrow are indicating the break in the Coulomb peak of the SHT while the white (dashed) double arrow the distance between two SHT Coulomb peaks.}
\end{figure}

The Ge HWs used in this study were grown via the Stranski-Krastanow (SK) growth mechanism \cite{Stangl2004, Zhang2012}. 6.6\,\AA \,of Ge was deposited on a Si buffer layer, leading to the formation of hut clusters \cite{Mo1990}. After a subsequent annealing process of roughly 3\,h, in-plane Ge HWs with lengths of up to 1\,$\mu$m were achieved. In the last step of the growth process, the wires were covered with a 3-5\,nm thick Si cap to prevent oxidation of the Ge \cite{Watzinger2016}. HWs have well defined triangular cross-sections with an average base width of 18.6\,nm \cite{Zhang2012} and they are oriented solely along the [100] and the [010] directions. Their density is directly related to the amount of deposited Ge. For this study samples with a relatively high amount of Ge were used, resulting in a high density of HWs. This leads to an increased probability of 'collisions' between perpendicularly grown HWs. Due to the short-range strain repulsion \cite{Johnson:JAP}, the Ge HWs tend not to merge into each other but T-like two-wire structures can emerge on the Si substrate (Figure \ref{fig:figure1} (a)). The distance between the two HWs in such a T-like structure can be shorter than 10\,nm. For such distances, tunneling between P donors implanted in Si has been observed \cite{MorelloNature2010, Mahapatra2011}. Indeed, tunneling facilitated by the leakage of the hole wave function in the SiGe substrate \cite{Yakimov:SST} is observed also between two HWs, as will be shown below.

The devices studied in this work were fabricated out of the above mentioned T-like structures. Figure \ref{fig:figure1} (b) shows a schematic of a device used for the charge sensing experiment. The upper wire, contacted with source, drain and gate electrodes, acts as a single hole transistor (SHT) and it is used both as a charge sensor and as a reservoir for holes. In the other wire, contacted only with a gate electrode, we create a QD that can host a spin qubit. This QD is formed presumably between the gate and the end of the wire; its hole occupation is to be determined with the SHT sensor coupled to it. The metal electrodes were defined by means of electron beam lithography. For the source and the drain electrodes, a few tens (20-40\,nm) of Pd, Pt or combination of Pd/Al contacts were used. Before the metal deposition, a short dip in buffered hydrofluoric acid is performed in order to remove the native oxide. 6-8\,nm  of hafnium oxide deposited by atomic layer deposition act as an insulator between the HWs with source and drain electrodes and the  Ti/Pd (5/20\,nm) side gate electrodes, defined in the last step of fabrication. 
A scanning electron micrograph of a typical device is shown in Figure \ref{fig:figure1}~(c). For two coupled QDs, the electrochemical potential of one QD depends on the charge state of the other \cite{DQDReview}. This can be seen in Figure \ref{fig:figure1}~(d), where the ladder of electrochemical potentials of the sensor is illustrated for two different QD charge configurations, M and M+1 \cite{MorelloArchitecture}. Every time the condition for hole tunneling from the QD to the SHT is satisfied, this tunneling event will leave the dot with less holes, which will thus shift the electrochemical potentials of the SHT causing a break in the SHT Coulomb peak. In order to reach again the same SHT Coulomb peak the gate voltage of the sensor needs to be adapted. A stability diagram with characteristic breaks is shown in Figure \ref{fig:figure1}~(e). The charge-transfer signal, $\Delta q/e$, where 1\,e is equivalent to the distance between two adjacent sensor Coulomb peaks, is very pronounced and equal to 30$\%$ (see Figure \ref{fig:figure1}~(f)) and observable thus even at a temperature of 1.5\,K. We have measured charge sensing in four different T-like devices.

\begin{figure}[hhhhhhh!!!]
\includegraphics{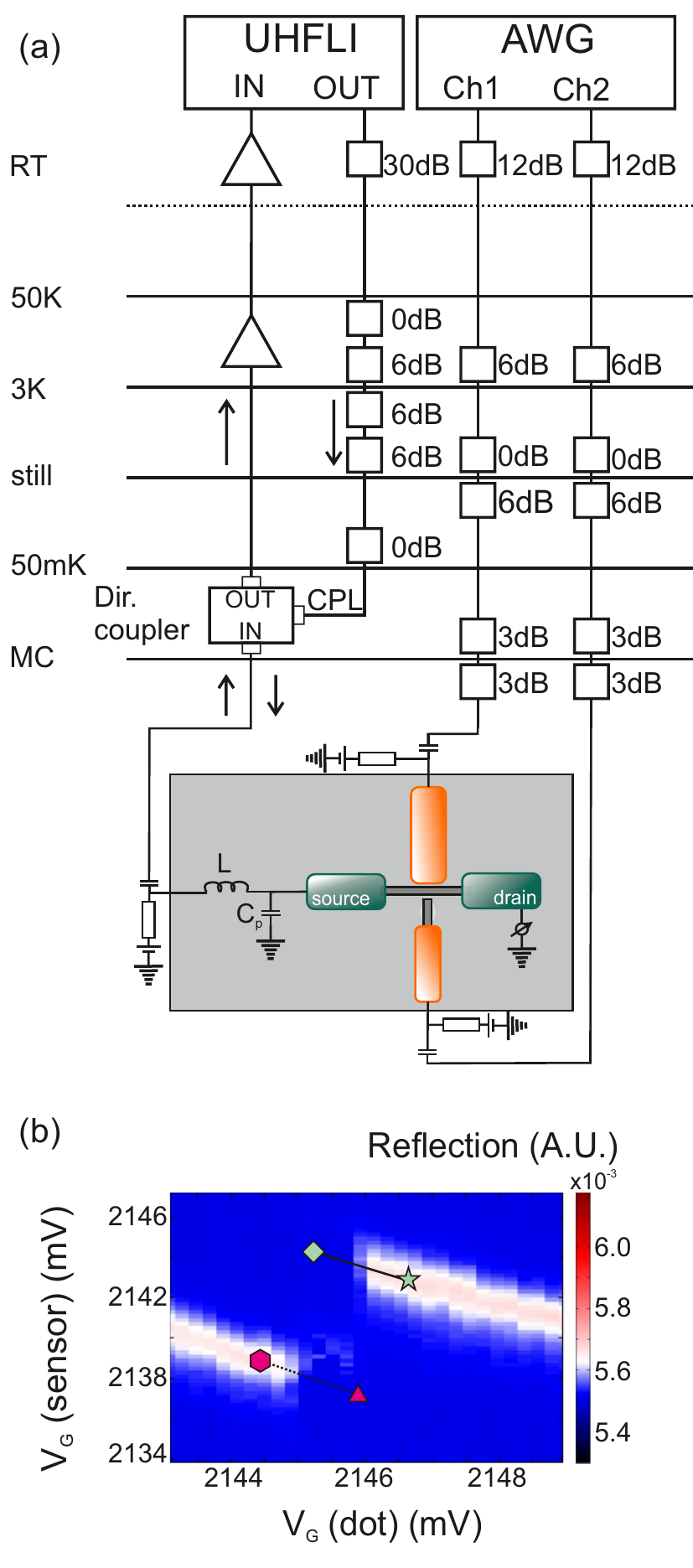}
\caption{\label{fig:figure2} (a) Simplified measurement circuit scheme. The source of the charge sensor is connected to the matching circuit formed with an inductor $L$ and the parasitic capacitance $C_p$ to ground. The RF signal is sent to the sample from an ultra-high frequency \textit{lock-in} (UHFLI) amplifier; it is attenuated at various dilution refrigerator stages. The reflected signal is amplified on two stages before the readout. Both gates of the device are connected to an AWG, which is used for applying short voltage pulses. (b) Zoom-in of a stability diagram, measured in reflection at a temperature of 30\,mK. The power of the RF signal on the \textit{lock-in} output was -35\,dBm, the low-pass filter bandwidth 100\,Hz and V$_{\textrm{SD}}$ 80\,$\mu$V. A green rhombus (star) and a pink hexagon (triangle) indicate the loading (unloading) position in the pulsing experiments and solid and dashed black lines the direction of pulsing.}
\end{figure}

In order to investigate whether the realized charge sensor is suitable for spin readout experiments, it was integrated into a resonant RF circuit and a reflectometry readout was performed. Additionally, the gates of the devices were connected to an arbitrary waveform generator (AWG), allowing fast gating. The resonant RF circuit consisted of a 2200\,nH inductor and the parasitic capacitance to the ground.  From the measured resonant frequency of 114.5\,MHz a parasitic capacitance of $\approx$0.9\,pF could be extracted. The higher measurement bandwidth due to the diminished 1/f noise and the sensitivity to both capacitive and resistive changes of the device are the main advantages of using this type of readout technique. The scheme for the RF reflectometry and fast gating setup is shown in Figure \ref{fig:figure2}~(a). Figure \ref{fig:figure2}~(b) shows a zoom-in into a stability diagram of a second device similar to that shown in Figure \ref{fig:figure1}~(e). In contrast to Figure \ref{fig:figure1}~(e), here the measured quantity is the amplitude of the reflected RF signal, integrated over approximately 10\,ms and the measurement was performed in a dilution refrigerator at a base temperature of about 30\,mK.

\begin{figure}[hhhhhhh!!!]
\includegraphics{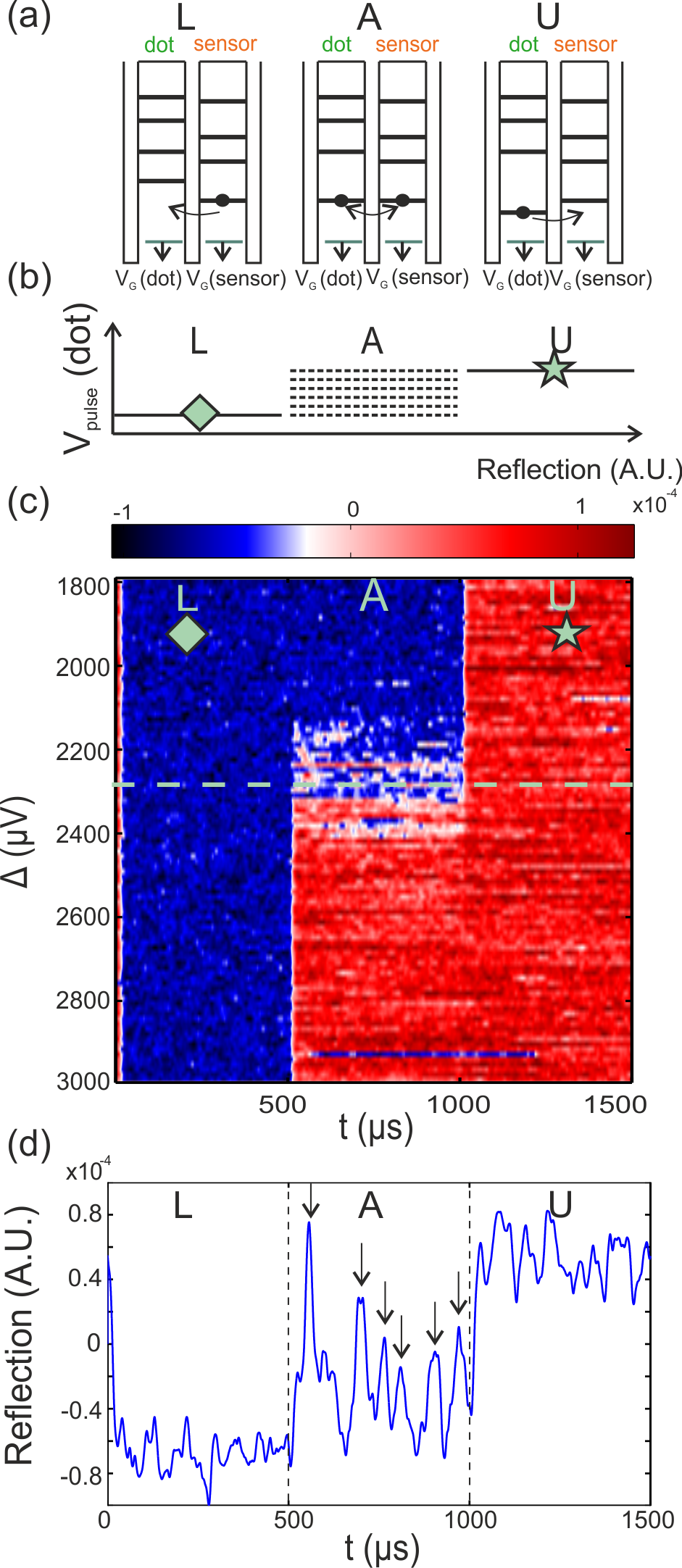}%
\caption{\label{fig:flagplot} (a) Scheme showing the alignment of the electrochemical potentials of the dot and the sensor for three different conditions. In the left part a hole is loaded in the dot (L); in the middle the resonant tunnelling condition is achieved by aligning the electrochemical potentials (A) and on the right side a hole is unloaded from the dot (U). (b) Schematic showing the shape of the applied three-part pulse. (c) Reflection amplitude of the sensor versus the relative voltage applied to the dot gate $\Delta$ in the align phase (A) and time $t$. The zero gate value corresponds to the load voltage. Loading and unloading of the hole is labeled with a green rhombus and a green star, respectively. (d) Single shot reflectometry trace corresponding to the position of the green dashed line in (c), where the condition for resonant tunneling is met. In the second part of the pulse ($500\mu s<t<1000 \mu s$) several hole tunneling events can be observed, indicated by black arrows. The power of the RF signal on the \textit{lock-in} output was -35\,dBm, the low-pass filter bandwidth 20\,kHz and V$_{\textrm{SD}}$ 80\,$\mu$V.}
\end{figure}

For real time detection of tunneling events between the QD and the sensor fast pulsing was used; a three-part voltage pulse was applied to the gates of the device. The pulse was applied along the upper part of the break in the Coulomb peak of the SHT shown in Figure \ref{fig:figure2}~(b) (black solid line). This diagonal pulsing is achieved by applying the pulse both to the dot and to the sensor gate simultaneously, but with a different sign and with a different amplitude. Each part of the pulse lasted for 500\,$\mu$s. With the first part of the pulse a hole is loaded into the dot (left part in Figure \ref{fig:flagplot}~(a); green rhombus in Figures \ref{fig:flagplot}~(b) and (c)), and with the last part a hole is unloaded (right part in Figure \ref{fig:flagplot}~(a), green star in Figures \ref{fig:flagplot}~(b) and (c)); the reflection amplitude shows a minimum value when the hole is loaded into the QD and a maximum value when it is localized in the SHT. In between those two parts of pulses an additional one is applied aiming to align the electrochemical potentials of the QD and the SHT (middle part in Figure \ref{fig:flagplot}~(a)). The voltage amplitude of this middle part of the pulse was varied in each of the 100 pulses which were applied. The schematic of the applied pulse is shown in Figure \ref{fig:flagplot}~(b) and the reflected signal from the sensor in Figure \ref{fig:flagplot}~(c). When the electrochemical potentials between the QD and the SHT are aligned, continuous exchange of holes between the QD and the sensor can take place. This can be indeed observed in Figure \ref{fig:flagplot}~(c) for dot gate voltage levels between 2127$\mu$V and 2418 $\mu$V.  The line trace shown in Figure \ref{fig:flagplot}~(d), taken at the position of the green dashed line in Figure \ref{fig:flagplot}~(c), shows indeed several tunneling events during the align pulse time. The small and unequal peak heights of the tunneling events are due to the limited bandwidth of the used setup.

The fact that we are performing transport measurements in the growth plane of the HWs, which have been shown to host HH state\cite{Watzinger2016}, implies that the HH mass in that direction is $m^*_{\textrm{HH}}\approx m/(\gamma_1 + \gamma_2)$, where $\gamma_1$ and $\gamma_2$ are Luttinger parameters \cite{KatsarosPRL2011}. This leads to a $m^*_{\textrm{HH}}$ of about 0.057 for Ge. Since this effective mass is smaller than the effective mass of electrons in Si, we expect shorter tunneling times than those reported for electrons (from 100\,$\mu$s to 10\,ms range) \cite{Mahapatra2011, Dzurak2010, PlaNature2012}.

\begin{figure}[hhhhhhhhhhhhhhhhhh!!!!!!!!!!!]
\center
\includegraphics{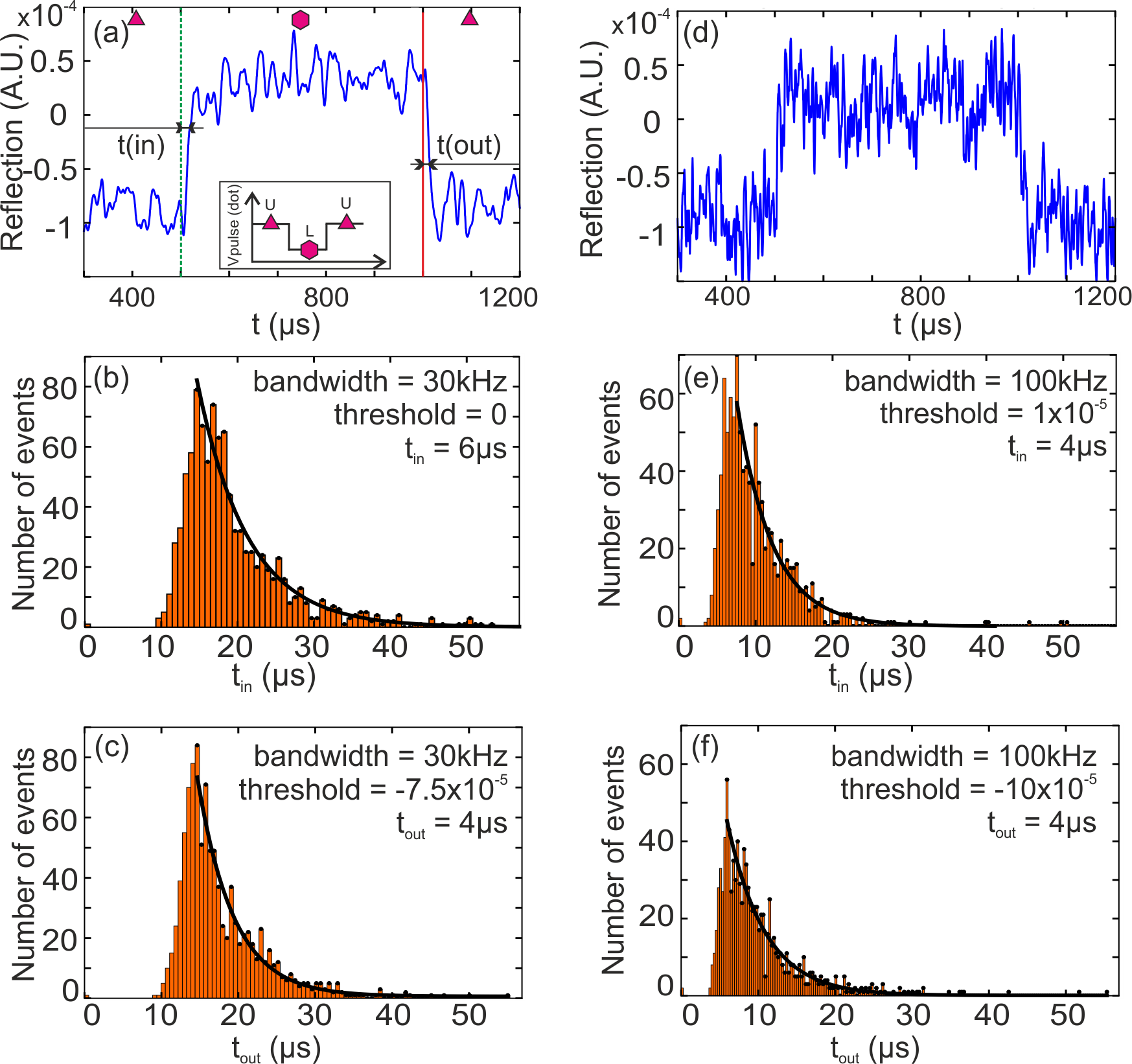}

\caption{\label{fig:figure4} (a) Single-shot reflection amplitude measurement of the sensor vs time, taken at the lower part of the break in the Coulomb peak in Figure \ref{fig:figure2}~(b), with the bandwidth of 30\,kHz. The reflection amplitude is recorded while unloading (labeled by triangles), loading (labeled by a hexagon) and again unloading a hole from the QD. The reflection amplitude is negative due to the offset in the instrumentation. The green dashed (red solid) line indicates the time when the loading (unloading) pulse was applied. The horizontal black lines indicate a threshold value above (below) which a tunneling in or tunneling out event is considered to have happened. The inset shows the shape of the applied pulse. (d) Single-shot reflection amplitude measurement of the sensor vs time, taken with the bandwidth of 100\,kHz. (b)-(c) and (e)-(f) Histograms of the delay times for loading and unloading the dot for 30\,kHz and 100\,kHz, respectively. From the exponential fit (solid black line) the tunneling times were extracted. The counts at the beginning of the histograms are attributed to the Gaussian noise distribution.}
\end{figure}

As already indicated above, due to the limited setup bandwidth, the extraction of the hole tunneling times cannot be achieved from an experiment similar to that described in Figure \ref{fig:flagplot}. In order to circumvent the problem of the slow rise time an experiment was devised in which the information of a tunneling event was encoded in a signal of a much longer duration than the rise time (see the Supporting Information). A three-part voltage pulse was now applied along the lower part of the break in the Coulomb peak in Figure \ref{fig:figure2}~(b) (black dashed line) in order to load (pink hexagon in Figure \ref{fig:figure2}~(b)) and unload (pink triangle in Figure \ref{fig:figure2}~(b)) a hole into/from the dot. Again each part of the pulse lasted  for 500\,$\mu$s. The shape of the applied pulse is shown in the inset in Figure \ref{fig:figure4}~(a), with pink triangles labeling the position when a hole is unloaded from the QD and a pink hexagon labeling the position when a hole is loaded into the dot. 

A 30\,kHz bandwidth single-shot reflection amplitude measurement of the sensor during the three-part pulse is shown in Figure \ref{fig:figure4}~(a). During the first part of the applied voltage pulse, when a hole is removed from the dot (pink triangle in Figure \ref{fig:figure2}~(b)), the reflected signal is at its minimum. With the second, negative voltage pulse part, a hole is loaded into the dot (pink hexagon in Figure \ref{fig:figure2}~(b)); the reflected signal from the sensor reaches its maximum. Finally, the hole is again removed from the dot in the last stage of the pulse and the reflected signal returns to its minimum. The green dashed (red solid) line indicates the starting edge of the second (third) part of the pulse. 1000 such measurements were performed and the delay times (t$_{\textrm{in}}$ and t$_{\textrm{out}}$) were extracted (see Figure \ref{fig:figure4}~(a)). A hole was considered to have tunneled into (out of) the QD if the reflected signal was higher (lower) than a certain threshold value of the reflection amplitude. The extracted times are shown in the histogram plots, in Figure \ref{fig:figure4}~(b) for tunneling into the dot and in Figure \ref{fig:figure4}~(c) for tunneling out of the dot. From the exponential fit a tunneling-in time of 6\,$\mu$s  and a tunneling-out time of 4\,$\mu$s was determined for thresholds equal to 0 and $-7.5 \times 10^{-5}$, respectively. Different thresholds are chosen in order to largely avoid false counts coming from noise peaks surpassing the threshold value. It was set as high as possible (without reaching the average value of the load phase) for the tunnel-in and as low as possible (without reaching the average value of the unload phase) for the tunnel-out times. The tunneling times depend slightly on the chosen threshold, but are always between 2-10\,$\mu$s. The same measurement and analysis were repeated for the bandwidth of 100\,kHz (Figure \ref{fig:figure4} (e)-(g)). The extracted tunneling in and out times of 5\,$\mu$s for bandwidth of 100\,kHz do not differ from those extracted for the bandwidth of 30\,kHz within our experimental error. In all experiments no difference between tunneling in and tunneling out times could be observed. It is important to note that the extracted tunneling times are two to three orders of magnitude shorter than the predicted spin relaxation times \cite{Zinoveva2005}, allowing future single-shot spin readout experiments.

In conclusion, we have demonstrated charge sensing in Ge HWs based on a capacitive and tunnel coupling mechanism between a QD, to act as a host for a qubit, and an SHT. Successful implementation of RF reflectometry measurements enabled the detection of single-hole tunneling events. The observed large charge transfer signals and the extracted hole tunneling times of a few $\mu$s pave the way towards projective spin readout measurements. While our experiment is a first step towards a spin-to-charge conversion setup it is clear that in order to realize scalable architectures, growth on pre-patterned substrates will be needed. Such growth has been intensively investigated \cite{SchmidtBook, KatsarosPRL2008} in the past and successfully demonstrated for dome islands \cite{ZhongAPL2004, ZhangAPL2007}. Once the positioning of hut wires will be well controlled the realization of more complex devices, allowing thus the coupling of multiple qubits, will become possible.

The work was financially supported by the ERC Starting Grant no. 335497 project and the FWF-Y 715-N30 project. We acknowledge Thomas Watson for helpful discussions and F.~Sch\"affler for fruitfull discussions related to the HW growth and for giving us access to the molecular beam epitaxy system. This research was supported by the Scientific Service Units (SSU) of IST Austria through resources provided by the MIBA Machine Shop. In particular we acknowledge T. Asenov, T. Adeltzberger, T. Menner and P. Traunm\"uller. 

\textbf{Supporting Information}. \newline Discussion of the effect of the finite rise time of the low pass filter on the extracted tunneling times.

\newpage

\bibliography{LV_bibliography}

\end{document}